# High- and low-conductance NMDA receptors are present in layer 4 spiny stellate and layer 2/3 pyramidal neurons of mouse barrel cortex

**Christian Scheppach**[1,2]*

[1]Physiological Laboratory, Department of Physiology, Development and Neuroscience, University of Cambridge, Cambridge, U.K.

[2]Institute of Physics, University of Freiburg, Freiburg, Germany

**\* Correspondence:** Christian Scheppach, Institute of Physics, University of Freiburg, Hermann-Herder-Str. 3, Freiburg, D-79104, Germany.
christian.scheppach@physik.uni-freiburg.de

**Abstract:** NMDA receptors are ion channels activated by the neurotransmitter glutamate in the mammalian brain and are important in synaptic function and plasticity, but are also found in extrasynaptic locations and influence neuronal excitability. There are different NMDA receptor subtypes which differ in their single-channel conductance. Recently, synaptic plasticity has been studied in mouse barrel cortex, the primary sensory cortex for input from the animal's whiskers. Pharmacological data imply the presence of low-conductance NMDA receptors in spiny stellate neurons of cortical layer 4, but of high-conductance NMDA receptors in pyramidal neurons of layer 2/3. Here, to obtain complementary electrophysiological information on the functional NMDA receptors expressed in layer 4 and layer 2/3 neurons, single NMDA receptor currents were recorded with the patch-clamp method. Both cell types were found to contain high-conductance as well as low-conductance NMDA receptors. The results are consistent with the reported pharmacological data on synaptic plasticity, and with previous claims of a prominent role of low-conductance NMDA receptors in layer 4 spiny stellate neurons, including broad integration, amplification and distribution of excitation within the barrel in response to whisker stimulation, as well as modulation of excitability by ambient glutamate. However, layer 4 cells also expressed high-conductance NMDA receptors. The presence of low-conductance NMDA receptors in layer 2/3 pyramidal neurons suggests that some of these functions may be shared with layer 4 spiny stellate neurons.

**Abbreviations:** EPSC, excitatory postsynaptic current; GluN2AR, GluN2A receptor, i.e. NMDAR with two GluN2A subunits - likewise for GluN2B, C, D; IV curve, current-voltage curve; L2/3, layer 2/3; L4, layer 4; NMDA, N-methyl-D-aspartate; NMDAR, NMDA receptor; RP, resting potential; t-LTD, timing-dependent long-term depression; t-LTP, timing-dependent long-term potentiation.

**Funding information:** This work was funded by the German Academic Exchange Service (DAAD), the Biotechnology and Biological Sciences Research Council (BBSRC), the Cambridge European Trust (CET) and the Research Innovation Fund of the University of Freiburg.

# 1  Introduction

N-Methyl-D-aspartate (NMDA) receptors constitute a major class of glutamate receptors in the mammalian brain (Traynelis et al., 2010). They contribute to the excitatory postsynaptic current (EPSC, Bekkers and Stevens, 1989) and are crucial in synaptic plasticity (Citri and Malenka, 2008), but also subserve other neuronal processes, for example dendritic NMDA spikes (Schiller et al., 2000; review: Major et al., 2013) or the sensing of ambient (Sah et al., 1989) and synaptic spill-over glutamate (Kullmann et al., 1996), and they can be present in synaptic and extrasynaptic locations (Stocca and Vicini, 1998; Thomas et al., 2006; Hardingham and Bading, 2010).

A number of NMDA receptor (NMDAR) subtypes are known, with different properties (Cull-Candy and Leszkiewicz, 2004), presumably adapted to the roles they play in different systems. NMDARs are tetramers, typically consisting of two GluN1 and two GluN2 subunits. The four known GluN2 subunits A, B, C and D endow the receptor with distinctive properties. For brevity, one can refer to NMDARs containing two GluN2A subunits as GluN2A receptors (GluN2ARs), and likewise for B, C and D. GluN2A and GluN2B receptors have a higher single channel conductance than GluN2C or GluN2D receptors. NMDARs with conductances in the higher range are therefore called high-conductance NMDARs, and if their conductances are in the lower range, they are called low-conductance NMDARs. Next to these diheteromeric NMDARs with two GluN1 and two identical GluN2 subunits, triheteromeric assemblies have been described, for example with one GluN2A and one GluN2B subunit (Paoletti et al., 2013), and there are two further subunit types, GluN3A and GluN3B (Low and Wee, 2010; Pachernegg et al., 2012), activated by glycine.

Barrel cortex (Fox, 2008), the primary sensory cortex of the whiskers of rodents, is one of the most intensively-studied regions of the mammalian neocortex. Sensory input from the whiskers is received via the thalamus, and the thalamo-cortical input fibers project to spiny stellate neurons of layer 4 (L4) of barrel cortex. L4 spiny stellate cells make vertical connections to pyramidal neurons of layer 2/3 (L2/3), usually within the same barrel. These cells, in turn, project horizontally to other L2/3 pyramidal neurons, within and across barrels (see Fig. 1).

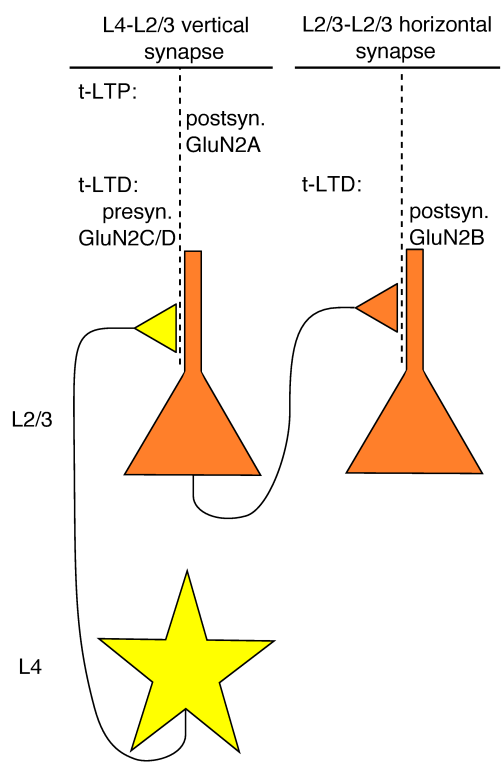

**Figure 1: NMDA receptor involvement in plasticity at synapses of barrel cortex.** At vertical synapses from L4 spiny stellate neurons to L2/3 pyramidal neurons, t-LTP requires postsynaptic GluN2A-containing NMDARs, whereas t-LTD requires presynaptic GluN2C or -D-containing receptors. In contrast, t-LTD at horizontal synapses between L2/3 pyramidal cells requires postsynaptic GluN2B-containing receptors.

NMDARs are crucial to synaptic plasticity both at the L4-L2/3 vertical synapse and the L2/3-L2/3 horizontal synapse, but the plasticity mechanisms, the NMDAR subtypes involved and their location seem to be different at these two synapses and also to depend on the type of plasticity. When spike timing dependent plasticity (Markram et al., 1997) is studied at the L4-L2/3 synapse, timing-dependent long-term potentiation (t-LTP) requires postsynaptic but not presynaptic NMDARs, whereas timing-dependent long-term depression (t-LTD) requires presynaptic but not postsynaptic NMDARs (Rodriguez-Moreno and Paulsen, 2008). In contrast, t-LTD at the L2/3-L2/3 synapse requires postsynaptic but not presynaptic NMDARs (Banerjee et al., 2014). Furthermore, the subunit composition of the NMDARs involved has been studied with pharmacological blockers (Banerjee et al., 2009; review: Rodriguez-Moreno et al., 2010). The presynaptic NMDARs involved in t-LTD at the L4-L2/3 synapse may contain GluN2C or GluN2D subunits, whereas the postsynaptic NMDARs involved in t-LTP may contain the GluN2A subunit. In contrast, postsynaptic GluN2B receptors seem to be necessary for t-LTD at the L2/3-L2/3 synapse (see Fig. 1).

Next to their role in synaptic plasticity, NMDARs have other important roles in L4 spiny stellate cells of barrel cortex. Apart from receiving the thalamic input, these cells also form strong excitatory connections to other L4 spiny stellate cells of the same barrel, thereby amplifying and distributing the afferent thalamic activity within the barrel (Feldmeyer et al., 1999). The EPSC has a relatively large NMDAR component, even at potentials around rest, which has been attributed either to a large number of GluN2A- or GluN2B-containing NMDARs at the synapse (Feldmeyer et al., 1999) or to the presence of GluN2C-containing NMDARs, which are less susceptible to block by $Mg^{2+}$ (Fleidervish et al., 1998; Binshtok et al., 2006). Furthermore, tonic NMDAR stimulation by ambient glutamate has been shown to influence the resting potential and, hence, excitability of L4 spiny stellate neurons (Binshtok et al., 2006). Finally, their dendrites exhibit NMDA spikes, which contribute to the angular tuning of L4 spiny stellate cells when responding to whisker deflections (Lavzin et al., 2012).

The proposed mechanisms of plasticity at the L4-L2/3 and the L2/3-L2/3 synapse are solely based on pharmacology (Rodriguez-Moreno and Paulsen, 2008; Banerjee et al., 2009, 2014). Hence, complementary electrophysiological information on the functional NMDARs expressed in L4 and L2/3 neurons would be useful. With patch-clamp recordings of single NMDAR channel currents, one could distinguish between NMDAR subtypes on the basis of single-channel conductance. In addition, this experiment would address the controversy about the prominent role of GluN2C-containing NMDARs in L4 spiny stellate cells (Feldmeyer et al., 1999; Fleidervish et al., 1998; Binshtok et al., 2006), and it could help resolve the question whether the presumed GluN2C-dependent mechanisms are unique to L4 cells.

From early histological expression studies of NMDAR subtypes in the brain (Monyer et al., 1994), one would generally expect to find GluN2A- or GluN2B-containing NMDARs in cortical excitatory neurons, the presence of other subtypes would be surprising. Pharmacological experiments on synaptic plasticity in barrel cortex (Rodriguez-Moreno and Paulsen, 2008; Banerjee et al., 2009, 2014) suggest the expression of low-conductance GluN2C- or GluN2D-containing NMDARs in L4 spiny stellate cells, but of high-conductance GluN2A- and GluN2B-containing NMDARs in L2/3 pyramidal neurons. Binshtok et al. (2006) corroborate the presence of GluN2C-containing receptors in L4 spiny stellates, reporting single-channel recordings and histological data on the expression of the GluN2C subunit (see also Suchanek et al., 1997).

In the present study, single NMDAR currents were recorded, and both high- and low-conductance NMDARs were found in L2/3 pyramidal and L4 spiny stellate neurons, which is compatible with proposed plasticity mechanisms, but suggests that L4 spiny stellate neurons show considerable expression of functional high-conductance NMDARs in addition to GluN2C, and that expression of low-conductance NMDARs is not unique to L4 in barrel cortex.

## 2  Methods

### 2.1  Animals, brain slices

Animal procedures were in accordance with guidelines of the University of Cambridge and U.K. Home Office legislation. Acute brain slices were obtained from C57BL6 mice aged 10-17 days. Animals were killed by dislocation of the neck, and thalamocortical slices (Agmon and Connors, 1991) were prepared. The brain was cut with a razor blade at an angle as described in Agmon and Connors (1991), and 400 μm thick slices were obtained with a vibratome (Leica VT1200S). Slices were incubated at 34°C for 30 min and then kept at room temperature.

### 2.2  Solutions, chemicals

Extracellular solution (used for slicing and perfusion of slices): 125 mM NaCl, 2.5 mM KCl, 2 mM $CaCl_2$, 1 mM $MgCl_2$, 1.25 mM $NaH_2PO_4$, 25 mM $NaHCO_3$, 25 mM glucose, 10 μM glycine (NMDAR co-agonist, see Traynelis et al., 2010), bubbled with carbogen gas (95% $O_2$, 5% $CO_2$), pH 7.4 (c.f. Sakmann and Neher, 1995, p. 200; Spruston et al., 1995; Vargas-Caballero and Robinson, 2004).

Extracellular pipette solution: 145 mM NaCl, 2.5 mM KCl, 2 mM $CaCl_2$, 10 μM glycine, 10 μM CNQX (AMPA and kainate receptor blocker; CNQX disodium salt was obtained from Tocris), 10 mM HEPES, pH was adjusted to 7.4 with about 4.2 mM NaOH. (Mg-free to prevent Mg-block of NMDARs, see Cull-Candy, 2007.)

L-glutamic acid was obtained from Sigma-Aldrich. N-methyl-D-aspartic acid was obtained from Tocris.

### 2.3  Pipettes

Patch pipettes were pulled from borosilicate glass capillaries (Harvard Apparatus, 1.5 mm outer diameter, 0.86 mm inner diameter, filamented glass; Narishige gravity puller, two-stage pull, first drop about 7 mm) and fire-polished. The pipette size was estimated by measuring the bubble number (Sakmann and Neher, 1983, p. 66f). Bubble numbers were between 3.5 and 4.5, corresponding to tip resistances (with the extracellular pipette solution) between 18 and 14 MΩ.

### 2.4  Setup, recording

Slices were viewed under an Olympus BX50WI fixed stage upright microscope with a x60 objective, or a x10 objective for visualizing the barrels, with infrared (IR) or visible light differential interference contrast (DIC) optics and a camera. To assist orientation in the slice, the current position in the slicing plane was monitored with an optical position encoder (Renishaw). Cell-attached patch-clamp recordings were established, with typical seal resistances around 15 GΩ. Voltage clamp data were collected with a MultiClamp 700B (Axon Instruments) amplifier, with a feedback resistor of 50 GΩ and a 10 kHz 4-pole Bessel filter, and digitized at a sampling frequency of 50 kHz. The amplifier was controlled and the data acquired with a custom Matlab program written by H. Robinson. During the experiment, perfusion with extracellular solution was maintained in the recording chamber with a gravity-fed inlet and a suction outlet. All experiments were performed at room temperature (20-23°C).

## 2.5 Targeting L2/3 pyramidal neurons and L4 spiny stellate neurons of barrel cortex

In thalamocortical slices, barrel cortex is cut perpendicular to the brain surface, such that all six cortical layers are present in the slice, and the barrels at cortical layer 4 are visible under the microscope. Barrel cortex was visually identified under x10 magnification. Layer 4 was marked by the extent of the barrels. L2/3 pyramidal neurons and L4 spiny stellate neurons were identified by their location in the slice, as well as the distinctive shapes of their cell bodies. L2/3 pyramidal neurons are triangular-shaped and polar with a clear apical dendrite pointing towards the cortex surface, whereas L4 spiny stellate neurons appear apolar and spherical ("granular").

## 2.6 Stimuli

NMDAR currents were recorded in cell-attached mode in voltage-clamp, with the outside of the membrane held at constant voltages cycling through −30 mV, −20 mV, ..., 30 mV. Steps to these test voltages were 900 ms in duration, with 450 ms intervals at 0 mV between them.

## 2.7 Data analysis

Analysis was done in Matlab with a custom program written by C. Scheppach, with a database (SQLite) for book-keeping of the data and analysis results. All current traces were digitally low-pass filtered with a Gaussian filter at a cut-off frequency $f_c$ = 1 kHz. Leak subtraction of current traces was done manually, by identifying and specifying putative points of the baseline current, for example in stretches without channel activity, and interpolating between them (spline interpolation, or smoothing spline interpolation, see Matlab function csaps). Amplitude histograms of current traces were obtained by first subtracting the leak current, and then computing the histogram with a bin width of 0.01 pA. Gaussian functions were fitted to the histogram peaks with the Matlab function fminsearch (least squares fit).

## 2.8 Statistics

Statistical significance of differences in patch counts was tested with chi-squared tests ("contingency tables", see Howell, 2010). If the counts in two situations are compared, and there is one fitted parameter (the "hit probability"), one can obtain a $\chi^2$ value which is chi-squared distributed with 1 degree of freedom (d.o.f.). The integrated chi-squared distribution corresponding to the $\chi^2$ value yields the $p$-value. If $p$ was smaller than a significance level of α=5%, the difference of counts was called "statistically significant".

# 3 Result

## 3.1 Channel identification

Single channel recordings of NMDA receptors were obtained from L2/3 pyramidal neurons and L4 spiny stellate neurons of mouse barrel cortex. The aim was to obtain information about the NMDAR subtypes present in these cells. The strategy to differentiate between NMDAR subtypes was based on their difference in single channel conductance.

GluN2A and GluN2B receptors both have a main conductance level of 50 pS and a subconductance level of 40 pS (Stern et al., 1992; review: Cull-Candy and Leszkiewicz, 2004), whereas GluN2C receptors have conductance levels of 35 pS and 22 pS (Stern et al., 1992; review: Cull-Candy and Leszkiewicz, 2004), and GluN2D conduct at 35 pS and 16 pS (Wyllie et al., 1996; Momiyama et al., 1996; review: Cull-Candy and Leszkiewicz, 2004). These values hold for an extracellular calcium concentration of 1 mM, but the present experiments were performed at $[Ca^{2+}]_o = 2$ mM, and NMDAR single-channel currents decrease with increasing $[Ca^{2+}]_o$ (Ascher and Nowak, 1988), by about 10% when going from $[Ca^{2+}]_o = 1$ mM to 2 mM (Gibb and Colquhoun, 1992; Wyllie et al., 1996). Therefore, conductances $\geq 36$ pS are expected for high-conductance NMDARs, but $\leq 31$ pS for low-conductance NMDARs, allowing for a distinction between these two NMDAR classes.

For each neuron studied, a cell-attached patch-clamp recording was established. To activate NMDA receptors, glutamate or NMDA was added to the pipette solution. Current traces were collected with a voltage-clamp protocol holding the outside of the membrane at constant voltages cycling through −30 mV, −20 mV, ..., 30 mV, to obtain a sufficient number of single-channel openings at a range of voltages to be able to measure the slope conductance and reversal potential. Patches were usually stable for about 5 minutes before the noise level increased, and during this time, current traces were collected. Fig. 2 panels A, C and D show examples of the obtained traces, with well-resolved channel openings.

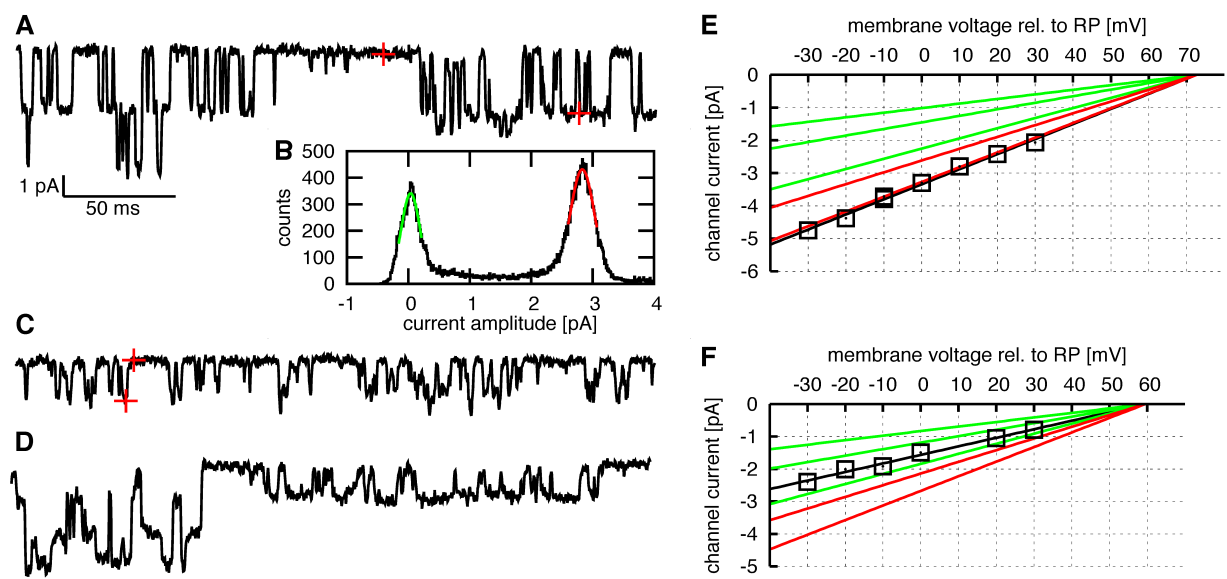

**Figure 2: High- and low-conductance NMDA receptor openings and IV-curves. (A)** Section of a current trace from a patch in which only high-conductance NMDARs were observed. Channel currents are negative (into the cell). Note the double openings in the first part of the trace. Red crosses: measurement points for the size of the single channel current, yielding 2.8 pA. (Data from a L2/3 pyramidal cell. Pipette glutamate concentration: 100 nM. Test voltage: 10 mV above resting potential (RP).) **(B)** Amplitude histogram of the current trace shown in (A). Gaussian functions were fitted to the baseline current peak (green) and the single opening peak (red), yielding the same single-channel current as obtained by cursor measurement in (A). **(C)** Section of a current trace from a patch in which only low-conductance NMDARs were observed. Red crosses: measurement points for the size of the single channel current, yielding 1.9 pA. (Data from a L4 spiny stellate cell. Pipette glutamate concentration: 75 nM. Test voltage: 10 mV below RP.) **(D)** Section of a current trace from a patch showing simultaneous activity from a high- and a low-conductance NMDAR. In the second half of the trace, only the low-conductance channel is active. In the first part, both the high- and the low-conductance channel are simultaneously active. All of the four possibilities (both closed, closed-open, open-closed, both open) and transitions between them can be observed. (Data from a L4 cell. Pipette glutamate concentration: 100 nM. Test voltage: RP − 10 mV.) **(E)** Current-voltage (IV) plot for the channel from which openings are shown in panel A. Black squares: measured single channel currents at a range of holding potentials from RP − 30 mV to RP + 30 mV. The data shown in panel A contribute the datapoint at RP + 10 mV. The total data come from a series of 8 sweeps, with 2 traces at RP − 10 mV and 1 trace for each of the other 6 voltages. Black line: straight line fit, yielding a slope conductance of 46.0 ± 1.5 pS and a reversal potential of RP + 73 mV. The red lines indicate slope conductances expected for the main and sub-conductance levels of GluN2A or GluN2B receptor channels (45 pS, 36 pS), the green lines for GluN2C and GluN2D receptors (31 pS, 20 pS, 14 pS), with the same reversal potential as the straight line fit to the experimental data (black line). On the basis of the slope conductance, this channel was identified as a high-conductance NMDAR. **(F)** IV plot for the channel from which openings are shown in panel C. Black squares: measured single channel currents at a range of holding potentials. The data shown in B contribute the datapoint at −10 mV. Black line: straight line fit, yielding a slope conductance of 26.3 ± 1.4 pS and a reversal potential of RP + 59 mV. Red and green lines as for (D). The observed slope conductance falls into the range for low-conductance NMDARs.

In analysis, for each patch, the available current traces were searched for single-channel openings. The single-channel currents were measured by visually-guided cursor selection (Fig. 2 A and C, red crosses) and collected in current-voltage (IV) plots (Fig. 2 E, F). Criteria to attribute openings in different traces to the same type of channel were 1) a consistent I-V relationship of the openings, 2) proximity in time of the traces in which the openings were observed and 3) similar further kinetic characteristics like typical opening time or burst duration. Only when such openings were frequent enough to be observed at a wide range of test voltages, was the channel further analyzed. Slope conductance and reversal potential of the putative channels were calculated from the IV plots by straight-line fits (Fig. 2 E, F).

NMDA receptors were identified by their expected linear current-voltage relationship for the chosen range of test voltages, a slope conductance between 14 pS and 45 pS and a reversal potential around 0 mV. Identification was also aided by factors like typical channel open durations in the order of 1-5 ms (Stern et al., 1992 and Wyllie et al., 1996 find mean open times between 0.6 and 3 ms for the different conductance levels of GluN2A, 2B, 2C and 2D receptors) and the expected pattern of subconductance levels, but these factors were not quantified.

Fig. 2 A shows a current trace from a L2/3 cell in which only high-conductance NMDARs were observed, and panel E shows the IV-curve corresponding to the channel openings seen in panel A. Fig. 2 C shows data from a L4 cell in which only low-conductance NMDARs could be seen, and panel F shows the corresponding single-channel IV-curve. In a number of patches, both channel types were present. Fig. 2 D shows a trace where a high- and a low-conductance NMDAR are active simultaneously. In about half of the patches with high-conductance NMDAR activity, multiple high-conductance openings could be observed (see Fig. 2 A, initial part of the trace), while the remainder seemed to contain only a single high-conductance NMDAR.

The range of test voltages (±30 mV around resting potential (RP)) and the pipette solution (glutamate or NMDA to activate NMDARs, AMPA and kainate receptors blocked by CNQX) were intended to make sure that mostly, NMDA receptors are the only channels active. These voltages are too hyperpolarized for recruitment of typical potassium channels; in addition, potassium channels would be distinguishable by producing an outward current, whereas NMDAR currents are inward. Sodium channels should either not activate or deactivate soon after the beginning of a depolarizing voltage step. The activity of chloride or calcium channels would still be possible, but can be distinguished by their different reversal potentials. As a control, 3 patches were measured without glutamate and CNQX in the pipette, none of which showed any channel activity. These considerations strengthen the claim that the observed channel activity was due to NMDA receptors.

To validate the cursor-by-eye method to measure single NMDAR channel currents (Fig. 2 A, C), a more formal method based on amplitude histograms (Fig. 2 B) was used for comparison. In two example traces, all-point amplitude histograms (see Sakmann and Neher, 1995, p. 527f) were computed (see methods). Gaussian functions were fitted to the histogram peaks, yielding center values for the peaks, and hence measurements of the single-channel current. The values obtained like this were in agreement with the cursor-by-eye values at an accuracy of 0.1 pA.

### 3.2 Overall patch statistics

Patches from L2/3 and L4 cells were obtained and sorted into four categories according to whether no channels, only high-conductance NMDARs, only low-conductance NMDARs, or both high- and low-conductance NMDARs could be identified in the patch. Most data were obtained at a pipette glutamate concentration of 100 nM (see table 1). Fig. 3 shows the data as percentages of patches with high- or low-conductance NMDAR activity. For example, of the 23 patches from L2/3 cells, 6 contained l.c. NMDARs, i.e. 26% (Fig. 3 A, right bar). 10 of the 23 patches, i.e. 43%, contained either high- or low-conductance NMDARs (Fig. 3 A, dashed line). Looking only at the patches with NMDAR activity, both in L2/3 and in L4 cells, all patches contained high-conductance NMDARs (Fig. 3 A and B, left bars). Low-conductance NMDARs were found in 8/9=89% of active L4 patches, but only in 6/10=60% of active L2/3 patches (Fig. 3 C). However, this difference was not statistically significant (chi-squared test, 1 d.o.f., $\chi^2$=2.04, $p$=15%).

**Table 1: Numbers of patches from L2/3 and L4 cells showing high-conductance (h.c.) and/or low-conductance (l.c.) NMDAR activity, with 100 nM glutamate as agonist.** $n_{tot}$: total number of analyzed patches from L2/3 and L4 cells. $n_-$: numbers of patches in which no channels were identified. $n_{h.c.}$: only high-conductance NMDARs identified. $n_{l.c.}$: only low-conductance NMDARs identified. $n_{both}$: both high- and low-conductance NMDARs identified.

| cell type | $n_{tot}$ | $n_-$ | $n_{h.c.}$ | $n_{l.c.}$ | $n_{both}$ |
|---|---|---|---|---|---|
| L2/3 | 23 | 13 | 4 | 0 | 6 |
| L4 | 14 | 5 | 1 | 0 | 8 |

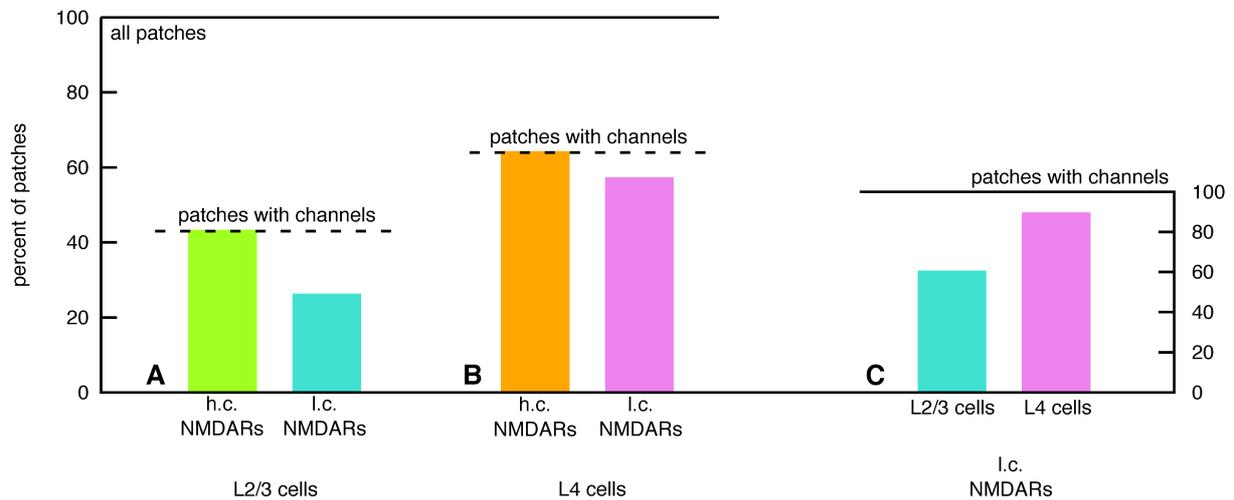

**Figure 3: Percentages of patches from L2/3 and L4 cells showing high-conductance (h.c.) and low-conductance (l.c.) NMDAR activity, in 100 nM glutamate. (A)** Percentages of patches from L2/3 cells showing h.c. and l.c. NMDAR activity. Dashed line: percentage of patches which showed NMDAR activity, irrespective of subtype. **(B)** the same for L4 cells. **(C)** compares the frequencies of l.c. NMDARs in L2/3 and L4 cells, when only active patches are taken into account.

For L4 cells, experiments were also performed with lower glutamate concentrations, down to 10 nM (see table 2). Fig. 4 shows the pooled data for all glutamate concentrations between 10 nM and 75 nM. The percentage of patches showing h.c. NMDAR activity was reduced compared to the data at 100 nM glutamate (chi-squared test, 1 d.o.f., $\chi^2$=4.68, $p$=3.1%), whereas the reduction of l.c. NMDAR activity was not statistically significant (chi-squared test, 1 d.o.f., $\chi^2$=0.536, $p$=46%). This is consistent with the higher glutamate sensitivities of GluN2C and GluN2D receptors (EC50 values for GluN2A and 2B receptors are 3.3 and 2.9 μM, but only 1.7 and 0.5 μM for GluN2C and 2D receptors, see (Erreger et al., 2007), corroborating the correctness of the identification of high- and low-conductance NMDARs.

**Table 2: Numbers of patches from L4 cells showing high- and low-conductance NMDAR activity, at glutamate concentrations below 100 nM.** $n_{tot}$: total number of analyzed patches at the given glutamate concentrations. $n_–$: numbers of patches in which no channels were identified. $n_{h.c.}$: only high-conductance NMDARs identified. $n_{l.c.}$: only low-conductance NMDARs identified. $n_{both}$: both high- and low-conductance NMDARs identified. The bottom row shows the patch counts pooled over the different glutamate concentrations.

| [Glu] | $n_{tot}$ | $n_–$ | $n_{h.c.}$ | $n_{l.c.}$ | $n_{both}$ |
|---|---|---|---|---|---|
| 75 nM | 4 | 2 | 0 | 1 | 1 |
| 50 nM | 9 | 6 | 1 | 1 | 1 |
| 10 nM | 3 | 0 | 0 | 2 | 1 |
| pooled | 16 | 8 | 1 | 4 | 3 |

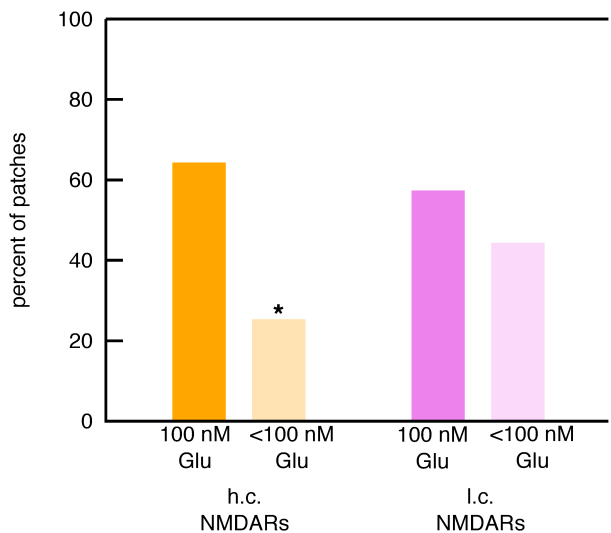

**Figure 4: Percentages of patches from L4 cells showing high- and low-conductance NMDAR activity, at glutamate concentrations below 100 nM.** The pooled data for glutamate concentrations between 10 nM and 75 nM are displayed (c.f. table 2). For

comparison, the percentages at 100 nM glutamate are also shown (c.f. Fig. 3 B). The reduction for h.c. NMDARs was statistically significant (*, $p$=3.1%).

As a control, experiments were also performed with 10 μM NMDA as agonist. Under these conditions, likewise, both high- and low-conductance NMDARs were observed in both L2/3 and L4 cells.

## 4  Discussion

### 4.1  Co-expression of high- and low-conductance NMDARs in L4 spiny stellate and L2/3 pyramidal neurons

NMDARs were studied in L4 spiny stellate and L2/3 pyramidal neurons of barrel cortex with cell-attached patch-clamp single-channel recordings, and NMDAR subtypes were distinguished by their single channel conductance. In both cell types, high-conductance NMDARs (e.g. GluN2A or GluN2B receptors) as well as low-conductance NMDARs (e.g. GluN2C or GluN2D receptors) were found. This co-expression, detected by single-channel recordings, with both NMDAR types frequently active in the same patch (Fig. 2 D), is analogous to experiments in granule cells of the cerebellum (Farrant et al., 1994), where expression of the GluN2C subunit is high (Monyer et al., 1992).

### 4.2  NMDARs in subcellular compartments of neurons

NMDAR recordings were obtained from somatic membrane, therefore the relationship to the NMDAR subtypes expressed at the presynapse, postsynapse and in the dendrites is only indirect. Pyramidal and spiny stellate cells generally do not receive excitatory synapses at the soma, but spines are limited to the dendrites, while the soma is targeted by inhibitory synapses. Hence, the NMDARs observed are unlikely to be of direct synaptic origin. NMDARs are synthesized in the somatic endoplasmic reticulum and trafficked along the dendrites to synaptic sites, where many NMDARs are not fixed but move between synaptic and extrasynaptic sites, diffusing along the membrane (Groc et al., 2009; Bard and Groc, 2011). From this dynamic picture of receptor localization (Choquet and Triller, 2013), if a cell expresses a certain NMDAR subtype in considerable quantity, one should be able to detect at least a small density at the soma as well.

### 4.3  Synaptic plasticity in barrel cortex

Pharmacological studies of spike-timing dependent plasticity (t-LTP and t-LTD) at the L4-L2/3 synapse and the L2/3-L2/3 synapse (Rodriguez-Moreno and Paulsen, 2008; Banerjee et al., 2009; Rodriguez-Moreno et al., 2010; Banerjee et al., 2014) suggest that L4 spiny stellate neurons express low-conductance GluN2C- or GluN2D-containing NMDARs, but L2/3 pyramidal neurons express high-conductance GluN2A- and GluN2B-containing NMDARs. Therefore the question arose whether these NMDAR subtypes can also be found electrophysiologically in the respective cell types. This could be confirmed (although high-conductance NMDARs were also found in L4 cells, and low-conductance NMDARs also in L2/3 cells), providing some complementary corroboration for the suggested plasticity mechanisms.

### 4.4  NMDARs in L4 spiny stellate neurons

In L4 spiny stellate cells, both high- and low-conductance NMDARs were observed, which is in agreement with Binshtok et al. (2006), although the present data suggest a rather higher density of high-conductance NMDARs. Out of 9 patches with channel activity (at 100 nM glutamate), 8 contained both high- and low-conductance NMDARs and 1 contained only high-conductance NMDARs (table 1, Fig. 3), whereas Binshtok et al. report 11 patches with only low-conductance NMDARs and 4 patches with both low- and high-conductance NMDARs, out of 15 patches showing channel activity. In the present study, at lower glutamate concentrations, the distribution appeared to shift in favor of low-conductance NMDARs (table 2, Fig. 4), indicating a possible agonist concentration effect.

The data are compatible with GluN2C-containing NMDARs at the postsynapse of connections between L4 spiny stellate neurons as well as at extrasynaptic sites (Binshtok et al., 2006), but also highlight the presence of high-conductance NMDARs in these cells. The thalamocortical synapses to L4 spiny stellate neurons are thought to have postsynaptic GluN2A- and GluN2B-containing NMDARs (Lu et al., 2001; Barth and Malenka, 2001), and spontaneous miniature EPSCs in L4 spiny stellate cells in thalamocortical slices have an NMDAR component which was found to be consistent with "canonical" GluN2A or GluN2B receptors (Espinosa and Kavalali, 2009).

### 4.5   NMDARs in L2/3 pyramidal neurons

In L2/3 pyramidal neurons of barrel cortex, likewise both high- and low-conductance NMDARs were found. To our knowledge, there are no previous reports of electrophysiological evidence for low-conductance NMDARs in this cell type. Generally, their presence in cortical excitatory neurons is unusual (histological expression study by Monyer et al., 1994) and contrasts for example with L5 pyramidal neurons of barrel cortex, where only high-conductance NMDARs were found (Binshtok et al., 2006). However, this is consistent with histological data reported by Binshtok et al., where expression of GluN2C was seen not only in L4 but also in L2/3, but no expression was seen in L5 pyramidal neurons.

Functional consequences of GluN2C-containing receptors have been discussed in L4 spiny stellate neurons. At synapses between these cells, the slow deactivation of GluN2C receptors and the resulting slow EPSC time-course would lead to strong and broad integration of inputs rather than sharp coincidence detection. Hence, the initial processing of sensory whisker input would be based on recurrent excitation, amplification and distribution of activity within the barrel (Feldmeyer et al., 1999; Fleidervish et al., 1998; Binshtok et al., 2006). Excessive GluN2C expression in L4 spiny stellate cells (and in L2/3 pyramidal neurons) has been linked to seizure generation in a mouse model of epilepsy (Lozovaya et al., 2014). Extrasynaptically, the lower sensitivity of GluN2C receptors to block by $Mg^{2+}$ and their higher glutamate sensitivity make them suitable to detect even low levels of ambient glutamate, even when the membrane potential is near rest, leading to changes of the resting potential and hence a modulation of neuronal excitability (Binshtok et al., 2006). The results of this study on low-conductance NMDARs being also present in L2/3 pyramidal neurons suggest that similar mechanisms may operate in these cells as well.

## 4.6 Dendritic NMDA spikes

Dendritic NMDA spikes have been reported in L4 spiny stellate neurons of barrel cortex (Lavzin et al., 2012) and in L2/3 pyramidal neurons of visual cortex (Smith et al., 2013) and somatosensory cortex (Palmer et al., 2014). Unlike in L5 pyramidal neurons, where GluN2A-containing receptors are thought to underlie dendritic NMDA spikes (Polsky et al., 2009), in L4 spiny stellate neurons, blocker experiments suggest that the current flows mainly through low-conductance NMDARs (Lavzin et al., 2012). One may suspect that the subtype-specific NMDAR kinetics influence the NMDA spike shape, but the spike duration appears similar in L5 pyramidal neurons (Schiller et al., 2000) and L4 spiny stellate neurons (Lavzin et al., 2012), so the NMDA spike shutdown may not be governed by NMDA receptor deactivation but by other conductances, for example repolarizing $K^+$ channels. Still, the low $Mg^{2+}$ block sensitivity and high glutamate sensitivity of low-conductance NMDARs may lead to a lower threshold for NMDA spike generation. The present electrophysiological results suggest that a mixture of high- and low-conductance NMDARs underlies dendritic NMDA spikes both in L4 spiny stellate and in L2/3 pyramidal neurons of barrel cortex.

## 4.7 Conclusion

In summary, the results provide complementary electrophysiological evidence for proposed synaptic plasticity mechanisms in barrel cortex, which were solely based on pharmacological findings thus far (section 4.3). The data are consistent with a prominent role of low-conductance NMDARs in L4 spiny stellate cells, but high-conductance NMDARs were equally present (section 4.4). Likewise, L2/3 pyramidal neurons of barrel cortex contained both high- and low-conductance NMDARs, suggesting that the mechanism of broad integration, amplification and distribution of excitation in response to sensory whisker input may apply not only to L4 but also to L2/3 neurons (section 4.5). The NMDAR subtype mix of high- and low-conductance NMDARs in the two cell types may be relevant to threshold properties of dendritic NMDA spikes (section 4.6).

## 5 Acknowledgements

I thank Hugh Robinson, Ole Paulsen and Abhishek Banerjee for their contributions to the conception and execution of the project, and Hugh Robinson and Ad Aertsen for critical reading and comments on the manuscript.

## 6 Conflict of interest

None.